# 500-period epitaxial Ge/Si$_{0.18}$Ge$_{0.82}$ multi-quantum wells on silicon

S. Assali,[1,*] S. Koelling,[1] Z. Abboud,[1] J. Nicolas,[1] A. Attiaoui,[1] and O. Moutanabbir[1]

[1] Department of Engineering Physics, École Polytechnique de Montréal, C. P. 6079, Succ. Centre-Ville, Montréal, Québec H3C 3A7, Canada

Ge/SiGe multi-quantum well heterostructures are highly sought-after for silicon-integrated optoelectronic devices operating in the broad range of the electromagnetic spectrum covering infrared to terahertz wavelengths. However, the epitaxial growth of these heterostructures at a thickness of a few microns has been a challenging task due the lattice mismatch and its associated instabilities resulting from the formation of growth defects. To elucidates these limits, we outline herein a process for the strain-balanced growth on silicon of 11.1 nm/21.5 nm Ge/Si$_{0.18}$Ge$_{0.82}$ superlattices (SLs) with a total thickness of 16 µm corresponding to 500 periods. Composition, thickness, and interface width are preserved across the entire SL heterostructure, which is an indication of limited Si-Ge intermixing. High crystallinity and low defect density are obtained in the Ge/Si$_{0.18}$Ge$_{0.82}$ layers, however, the dislocation pile up at the interface with the growth substrate induces micrometer-longs cracks on the surface. This eventually leads to significant layer tilt in the strain-balanced SL and in the formation of millimeter-long, free-standing flakes. These results confirm the local uniformity of structural properties and highlight the critical importance of threading dislocations in shaping the wafer-level stability of thick multi-quantum well heterostructures required to implement effective silicon-compatible Ge/SiGe photonic devices.

I. INTRODUCTION

Group IV Ge/SiGe heterostructures have been explored because of their potential as building blocks for monolithic, Si-integrated communication and sensing devices serving a broad range of the electromagnetic spectrum.[1] For instance, by carefully controlling the composition of Ge/Si$_{1-x}$Ge$_x$ multi-quantum well (MQW) heterostructures while preserving their high crystalline quality, the optical response can be precisely tailored for operation in near infrared through terahertz (THz).[2–6] Indeed, optical modulators, photodetectors, and waveguides at telecommunication wavelengths have been established using strain-balanced Ge/Si$_{0.16}$Ge$_{0.84}$ MQWs.[2,7] Photonic devices serving longer wavelengths are also at reach using this monolithic approach. Optical modulation in Si$_{0.8}$Ge$_{0.2}$ waveguides was recently demonstrated in the 5.5-11 μm range,[8] thus covering a large portion of the mid-wave and long-wave infrared.

Ge-rich heterostructures have also been proposed to implement active components in THz.[9] Sensing, spectroscopy, imaging,[10–12] astronomy,[13] and wireless[14] applications would all benefit from the availability of compact and manufacturable THz devices. With this perspective, strain-balanced n-type Ge/Si$_{0.8}$Ge$_{0.2}$ superlattice (SL) heterostructures have been proposed as a promising platform to implement room-temperature THz quantum cascade lasers (QCL).[9,15] This material system consists of a periodic stack of multiple Ge QWs that are separated by Si$_{0.8}$Ge$_{0.2}$ tunneling barriers. Under an applied bias, resonant tunneling of L-minimum electrons creates population inversion and result in THz emission via intersubband (ISB) transitions.[16,17] To achieve lasing emission, these devices require a thickness of the active region exceeding 10 μm, while keeping the Ge-Si$_{1-x}$Ge$_x$ interface roughness below 0.2 nm.[9,18] Preserving the strain-balanced growth of the Ge/Si$_{0.8}$Ge$_{0.2}$ stacking while limiting the nucleation of dislocations is essential to guarantee high material quality through the whole active region of the SL. Moreover, high reproducibility of the few nm-thick epilayers through the entire SL is a prerequisite to control the uniformity of their electronic structure. Current state-of-the-art Ge/Si$_{0.8}$Ge$_{0.2}$ heterostructures of a few tens of periods showed promising results for future THz device integration, with the demonstration of intense ISB absorption[17,19] and room-temperature LEDs.[20] However, the total SL thickness in these studies remains below 1 µm, which is about one order of magnitude thinner than the optimal thickness for lasing emission. Exploring and understanding the structural properties of SL heterostructures made of hundreds of periods would provide the essential information for the design of practical THz devices.

This work reports on the growth of strain-balanced Ge/Si$_{0.18}$Ge$_{0.82}$ SLs with a total thickness up to 16 µm (500 periods) and addresses their structural properties and stability. 11 nm/21 nm Ge/Si$_{0.18}$Ge$_{0.82}$ SLs were grown on a Si$_{0.115}$Ge$_{0.875}$/Ge/Si substrate. Composition and thickness of the layers remain constant throughout the whole thickness of the heterostructure, showing high crystallinity and low defect density in the SL. Atom probe tomography (APT) studies demonstrate a slight increase in the Ge-Si$_{0.18}$Ge$_{0.82}$ interfaces with increasing SL thickness, thus indicating limited Si-Ge intermixing even after the growth of 500 periods. X-ray diffraction (XRD) analysis demonstrates the high crystallinity of the 16 µm-thick SL. However, the pile up of dislocations at the interface with the Si$_{0.115}$Ge$_{0.875}$ substrate results in significant layer tilt in the SL and in the formation of micrometer-long cracks on the surface. Nevertheless, high compositional uniformity and crystallinity is obtained away from the cracks. The latter made the strain-balanced SL brittle, with the formation of millimeter-long, free-standing MQW flakes at the end of the growth during the cooldown to room-temperature.

## II. EXPERIMENTAL DETAILS

Ge/Si$_{0.18}$Ge$_{0.82}$ SLs were grown on 100 mm Si (100) wafers in a low-pressure chemical vapor deposition (CVD) reactor using H$_2$ as carrier gas. First, a 2.9 μm-thick Ge virtual substrate (VS) was grown using germane (GeH$_4$) precursor following a two-temperature step process (450/600 °C) and a post-growth thermal cyclic annealing (>800 °C). A 1.8 μm-thick Si$_{0.115}$Ge$_{0.885}$ layer was then grown for 45 minutes at 600 °C and at chamber pressure of 20 Torr using disilane (Si$_2$H$_6$) and GeH$_4$ precursors, with molar fractions of 4·10$^{-5}$ and 9·10$^{-4}$, respectively. The Si$_2$H$_6$ flow was then shutdown prior to the growth of the first Ge layer for 18 seconds at 600 °C. Next, the Ge/Si$_{0.18}$Ge$_{0.82}$ growth at a thickness of 11/21 nm was performed for 30/18 seconds at 600 °C by introducing Si$_2$H$_6$ (molar fraction of 7·10$^{-5}$). The GeH$_4$ flow was kept constant throughout the entire Ge/Si$_{0.18}$Ge$_{0.82}$ SL growth. SLs with 50 and 500 periods were grown following this protocol, corresponding to a growth time of 40 minutes and 400 minutes, respectively. APT measurements were performed using a LEAP 5000XS from Cameca with a laser producing picosecond pulses at 355 nm at a variable repetition rate of a few 100 kHz. The tip-shaped samples for APT were prepared using a FIB according to the procedure discussed in Ref.[21] in a FEI Helios FIB using a Ga ion beam (5-30 kV range). XRD measurements were performed using a Bruker Discover D8. A 3 bounces Ge(220) 2-crystals analyzer was placed in front of the XRD detector to reduce mosaicity during the (224) reciprocal space map measurements.

## III. RESULTS AND DISCUSSION

The low-resolution cross-sectional scanning transmission electron micrograph (STEM) acquired on the 500-period SL sample is shown in Fig. 1a. A total thickness of 16 µm is estimated for the Ge/Si$_{0.18}$Ge$_{0.82}$ SL grown on top of the 1.8 µm/2.9 µm Si$_{0.115}$Ge$_{0.885}$/Ge-VS. The periodic SL structure is visible in the TEM images acquired on top (Fig. 1b), middle (Fig. 1c), and bottom (Fig. 1d) regions of the stacking. High crystalline quality of the epilayers is visible across the whole stacking, which reaches an overall thickness of 20 µm on the Si wafer. No threading dislocations are observed at the TEM imaging scale, which is a first indication of a strain-balanced heterostructure. Threading dislocations with a density above 5×10$^7$ cm$^{-2}$ are present in the Ge-VS[22] and their amount will increase during the overgrowth of lattice-mismatched Si$_{0.115}$Ge$_{0.885}$. Dislocations are largely confined in the substrate at the Si$_{0.115}$Ge$_{0.885}$-Ge interface, while the Ge/Si$_{0.18}$Ge$_{0.82}$ SL exhibits a higher crystallinity where only a few growth defects can be identified across the whole SL (yellow arrows in Fig. 1b-d). Bending of pre-existing threading dislocations propagating from the Si$_{0.115}$Ge$_{0.885}$/Ge into the Ge/Si$_{0.18}$Ge$_{0.82}$ SL occurs during growth and contribute to local distortions in the lattice structure. Furthermore, reduction in the threading dislocation density might also takes place during the SL growth due to dislocation filtering resulting from the tensile-strained Si$_{0.18}$Ge$_{0.82}$ growth on Ge.[23]

The high homogeneity in the thickness and composition of the SL periods is demonstrated in Fig. 2 by comparing the top (periods #491-500, Fig. 2a) and bottom (periods #1-9, Fig. 2b) region of stacking. Strikingly, after the growth of a 16 µm-thick SL both layer thickness (*t*) and interface width (*w*) remain almost identical to the first few periods. To further investigate the compositional broadening, energy dispersive X-ray (EDX) measurements were performed (Supporting Information S1). The Si compositional profiles, estimated along the <100> growth direction of the SL, are plotted in Fig. 2c. The top and bottom EDX profiles show almost perfect spatial overlap, in agreement with the high reproducibility of the growth highlighted by the STEM images. Layer thickness and interface width are evaluated by fitting the EDX compositional profile using a sigmoidal function:

$$f(x) = A + \frac{B}{1+e^{-\frac{x_0 \pm x}{\tau}}} \quad (1)$$

where $A$ is a vertical offset parameter (Si content in the layer), $B$ is a scaling parameter (maximum Si content), $x_0$ is the inflection point of the curve, and the sign of $x$ results in an increasing or a decreasing function. The interface width $w$ is then estimated as $w = 4\tau$, while the layer thickness $t$ is evaluated as $t = x_0^+ - x_0^-$. The values extracted from the fit are displayed in Fig. 2d. Remarkably, the thickness of the Ge/Si$_{0.18}$Ge$_{0.82}$ layers remains constant at 11.1±0.3 nm/21.5±0.3 nm throughout the entire 500 periods of the stacking. Moreover, interface widths in the 4-6 nm range are obtained in both top and bottom regions of the stacking. Since the sample was maintained at 600 °C for more than 6 hours to complete the entire SL growth, our data show that Si-Ge interdiffusion remains limited which does not compromise the material quality.

To accurately estimate the composition and the interface roughness, atom probe tomography (APT) measurements were performed on a thinner 50-period SL sample. This sample shows identical quality as what obtained in the 500-period (Supplementary Material S2). The 3D APT reconstruction is displayed in Fig. 2e, while the Ge and Si compositional profiles extracted along the <100> growth direction are plotted in Fig. 2f (Supplementary Material S3). The layer thickness $t$ and interface width $w$ values obtained from fitting the Si compositional profile using Eq. (1). The nominal period thickness of 33±1 nm remains constant across the SL, with a constant

Si$_{0.18}$Ge$_{0.82}$ layer thickness of 21.5±0.4 nm. The Ge layer has a thickness of 11.2±0.4 nm, however, the measured stoichiometry is Si$_{0.02}$Ge$_{0.98}$ indicating a parasitic Si incorporation during growth. A smaller interface width is recorded for the Si$_{0.18}$Ge$_{0.82}$/Ge interface ($w_{down}$) compared to the Ge/Si$_{0.18}$Ge$_{0.82}$ ($w_{up}$) interfaces, with values of 1.6±0.2 nm and 3.4±0.2 nm, respectively. This interface width slightly decreases as the period number increases (Supplementary Material S3). Note that the higher accuracy on the spatial distribution of atoms of APT compared to EDX results in lower interface width values. As mentioned above (Fig. 2c), interdiffusion has limited impact on the growth, which yields a small broadening of the SL interfaces as the growth progresses at 600 °C. Moreover, the ~2 at.% parasitic Si incorporation in the Ge layer is likely due to the residual Si$_2$H$_6$ precursor in the CVD chamber/gas line during the off transient.

To further elucidate the structural properties at a larger scale, 2θ-ω X-ray diffraction spectroscopy (XRD) measurements were performed around the (004) XRD order, as shown in Fig. 3a. The peaks associated with Si$_{0.12}$Ge$_{0.88}$ and Ge-VS are detected at 66.5° and 66.06°, respectively. In the 50- and 500-period SLs, Pendellosung fringes develop as a result of the variation in the out-of-plane lattice parameter across the heterostructure,[24,25] thus indicating a coherent epitaxy. A full-width at half maximum (FWHM) of 0.03-0.04° and a constant angular spacing of ~0.33° is measured in the 50-period SL. The FWHM increases to 0.05-0.06° in the 500-period SL, while the peak spacing remains constant at ~0.33°. The XRD spectra were then fitted using a quasi-kinematical model (Supplementary Material S4).[24] In the 50-period SL, the estimated Ge/Si$_{0.18}$Ge$_{0.82}$ layer thickness from XRD (21.5/11 nm) closely match the values from APT (21.5/11.2 nm). Expectedly, the simulated peaks are sharper than the measured ones due to the presence of structural defects, which are not included in the XRD model.[24] Similar results are obtained in the 500-period SL sample. In both samples, the layer thicknesses closely match the values from EDX and APT measurements, however, the difference in the XRD peak width between the measured and the simulated curves becomes more pronounced in thicker SL (Supplementary Material S4). We highlight that the increase in the interface broadening observed in the top region of the SL (Fig. 2d) and layer tilt will contribute to the increase in FWHM that is visible in the 500-period SL.

The strain-balanced growth of SLs is demonstrated in the Reciprocal Space Mapping (RSM) measurements around the asymmetrical (224) XRD order in Fig. 3b-c. As a reference, the (224) RSM map for the Si$_{0.12}$Ge$_{0.88}$/Ge-VS/Si is displayed in Fig. 3d. Because of the difference in the thermal expansion coefficient between Ge and Si, a residual tensile strain $\varepsilon_\parallel$ of < 0.2 % is present in the Ge-VS. In the 50-period SL (Fig. 3c), the satellite peaks (SL-3:+3) share the same $q_x = -5.015\ nm^{-1}$ of the Si$_{0.12}$Ge$_{0.88}$/Ge-VS, as expected in a strain-balanced heterostructure. These SL peaks are equally spaced along $q_z$ by ~0.31 $nm^{-1}$. The pseudomorphic growth is preserved in the 500-period SL (Fig. 3b), with a small shift to higher $q_x$ values (dashed vertical line refers to Si$_{0.115}$Ge$_{0.875}$/Ge-VS). The $q_z$ peak spacing remains constant at ~0.31 $nm^{-1}$ in all SLs, thus confirming the coherent growth and the uniform layer thickness across the stacking, in agreement with the TEM-APT measurements (Figs. 1&2).

Interestingly, diagonal stripes are visible in XRD data (Fig. 3b-c), which could hint to the presence of a significant tilt in this sample. To elucidate this behavior, RSM measurements around the symmetrical (004) XRD order were acquired on the 50- and 500-period SLs, as displayed in Fig. 4a-b. In the 50-period SL (Fig. 5b) the satellite peaks are centered at $q_x = 0$, which indicates a negligible layer tilt between the (001) planes of the SL and the Si$_{0.115}$Ge$_{0.875}$/Ge substrate. Note that the vertical spacing between SL peaks is identical to what was obtained in the (224) RSM map (Fig. 3c). For the 500-period SL (Fig. 4a), the main satellite peaks remain visible, however, at smaller $q_z$ values a low intensity signal develops that extends across the range of the measured $q_x$. To better visualize this behavior, the plot of the (004) XRD intensity as a function of $q_x$ (fixed $q_z$ values) for the first satellite peak (SL +1) is shown in Fig. 4c. It is noticeable that by increasing the number of SL periods the FWHM of the (004) peak decreases from $9.7 \pm 0.2 \cdot 10^{-3}\ nm^{-1}$ to $6.1 \pm 0.1 \cdot 10^{-3}\ nm^{-1}$. This observation would suggest an increase in the spatial correlation of dislocations in the heterostructure, specifically at the interface between two lattice-mismatched layers.[26] As the total thickness of the SL increases, dislocations pile up in the lower region of the stacking, especially at the interface with $Si_{0.115}Ge_{0.875}$, could result in an ordered network of dislocations.[27,28] Indeed, the (004) RSM map for the $Si_{0.115}Ge_{0.875}$/Ge-VS/Si substrate already shows a broadening of the $Si_{0.115}Ge_{0.875}$ peak along the $q_x$ (Supplementary Material S5). Misfit dislocations at the $Si_{0.115}Ge_{0.875}$-Ge interface [26] and the nucleation of threading dislocations[29,30] is likely the main source of pile up of threading dislocations during Ge/$Si_{0.18}Ge_{0.82}$ SL growth on top.[30] This effect is clearly visible in the 500-period SL as a large deviation from $q_x = 0$ (Fig. 4a).[26]

Nonetheless, the large layer tilt can hardly be explained by dislocations pile up alone. The surface morphology of the 500-period SL is displayed in a representative 60×60 μm² AFM image in Fig. 4d, while 20×20 μm² maps for all samples are shown in Supplementary Material S6. Cracks oriented along the <110> and <1-10> directions are detected on the sample surface. Few cracks were already present in $Si_{0.12}Ge_{0.88}$/Ge and their length increases up to tens of microns with increasing number of periods of the SL, resulting in flat regions that are surrounded by cracks extending down to a depth of a few microns. Square array of cracks oriented along both <110> and <1-10> directions are commonly observed in SiGe and in hybrid III-V/SiGe systems.[27,31–35]

Cracks are induced by dislocations pile up, due to the reduced local growth rate in the proximity of arrested dislocations. This creates a rougher surface during growth, which will further block the glide of dislocations, inducing more pile up, and resulting in the additional nucleation of dislocations to promote the relaxation.[27,28] This process is exacerbated in the 500-period SL, where the higher cracks density eventually leads to the release of mm-long MQW flakes that are partially, or even entirely, detached from the surface. The scanning electron micrograph (SEM) in Fig. 4e and the optical micrograph in Fig 4f show examples of tilted regions and flakes, respectively. Remarkably, a very high spatial homogeneity of the <100> compositional profile was obtained in the 500-period SL (Fig. 2d). This observation, combined with the high crystallinity estimated from the RSM maps, suggests that flaking of the sample most likely occurs at the end of the growth during the cooldown to room-temperature. The creation of these flakes, consisting of rather flat, free-standing Ge/SiGe SLs (Supporting Information S7), is possibly triggered by thermal expansion mismatch that can be exacerbated by the grown SL. The higher defect density in the heterostructure will result in the material becoming brittle and most likely trigger the formation of Ge-rich flakes during the cooldown to room temperature.

## IV. CONCLUSIONS

In summary, we demonstrated the growth of a 500-period Ge/Si$_{0.18}$Ge$_{0.82}$ SLs at a total thickness reaching 16 μm on Si wafer using Si$_{0.115}$Ge$_{0.875}$/Ge-VS as interlayer. The uniformity of the composition and thickness (11.1/21.5 nm) of the SL layers are preserved throughout the entire heterostructure. Interface broadening in the bottom region of the SL only slightly increases with additional SL growth on top, showing limited Si-Ge intermixing even after more than six hours growth at 600 °C. High crystallinity and low defect density are demonstrated by combining TEM and XRD structural analysis. Negligible layer tilt is obtained in SLs up to 50 periods, however, by increasing the total thickness to 500 periods a tilt develops as a result of dislocations pile up and <110>-oriented cracks at the surface. This also leads to the formation of mm-long, free-standing flakes during the cooldown to room-temperature. These results show that the growth of Ge/Si$_{0.18}$Ge$_{0.82}$ SLs with a total thickness exceeding 10 μm preserves the structural properties of the material during growth. However, a careful optimization of high quality Si$_{0.12}$Ge$_{0.88}$/Ge-VS/Si buffer layer is needed to avoid dislocations pile up that compromises the structural stability of the SL at the end of the growth, during the cooldown to room temperature. These results provide a promising path to implement new designs for photonic devices requiring thick and highly uniform SLs such as THz QCLs and QW infrared photodetectors.

**FIGURES CAPTIONS**

**Figure 1.** (a) Overview STEM image and schematic drawing of the Ge/Si$_{0.18}$Ge$_{0.82}$ SL (500-period)/Si$_{0.115}$Ge$_{0.875}$/Ge-VS/Si heterostructure. (b-d) TEM images acquired in the top (b), middle (c), and bottom (d) regions of the 500-period SL.

**Figure 2.** (a-b) STEM images acquired in the top (a) and bottom (b) regions of the 500-period Ge/Si$_{0.18}$Ge$_{0.82}$ SL. (c) EDX compositional profile for the Si atoms acquired in the top and bottom regions of the SL. To allow for comparison, only 9 periods are shown, and a relative horizontal scale is used. (d) Interface width $w$ and layer thickness $t$ as a function of the SL period obtained from fitting (c) with Eq. (1). (e-f) 3D APT reconstruction (e) and extracted Si, Ge compositional

profiles (f) along the <100> growth direction. The fitted curve using Eq. (1) is shown with a dashed orange line.

**Figure 3.** (a) 2θ-ω scans around the (004) X-ray diffraction order acquired on the 50-500 periods Ge/Si$_{0.18}$Ge$_{0.82}$ SLs and on the Si$_{0.115}$Ge$_{0.875}$/Ge-VS. (b-d) RSM around the asymmetrical (224) reflection for the 500 (b) and 50 (c) periods Ge/Si$_{0.18}$Ge$_{0.82}$ SLs and for the Si$_{0.115}$Ge$_{0.875}$/Ge-VS (d).

**Figure 4.** (a-b) RSM around the symmetrical (004) reflection for the 500 (a) and 50 (b) periods Ge/Si$_{0.18}$Ge$_{0.82}$ SLs. (c) Plot of the (004) XRD intensity as a function of $q_x$ for a fixed $q_z$ estimated on the SL +1 peak of the 50 and 500 periods SLs. The profile for the tilted region in the 500 periods SL is also shown for comparison. (d) 60 μm × 60 μm AFM image for the 500-period SLs. (e) SEM image of the 500-period SL showing a tilted flat region surrounded by cracks. (f) Optical micrograph of the 500-period SL where mm-long flakes are visible.

**AUTHOR INFORMATION**

Corresponding Author:

*E-mail: simone.assali@polymtl.ca

Notes

The authors declare no competing financial interest.

## DATA AVAILABILITY

The data that support the findings of this study are available from the corresponding author upon reasonable request.

## SUPPLEMENTARY MATERIAL

See the supplementary material for additional details on the TEM, APT, and XRD characterization, the AFM images of all samples, and the SEM images of the flakes.

## ACKNOWLEDGEMENTS

The authors thanks J. Bouchard for the technical support with the CVD system, A. Kumar and M. Zöllner for the help with the XRD analysis. O.M. acknowledges support from NSERC Canada (Discovery, SPG, and CRD Grants), Canada Research Chairs, Canada Foundation for Innovation, Mitacs, PRIMA Québec, and Defence Canada (Innovation for Defence Excellence and Security, IDEaS).

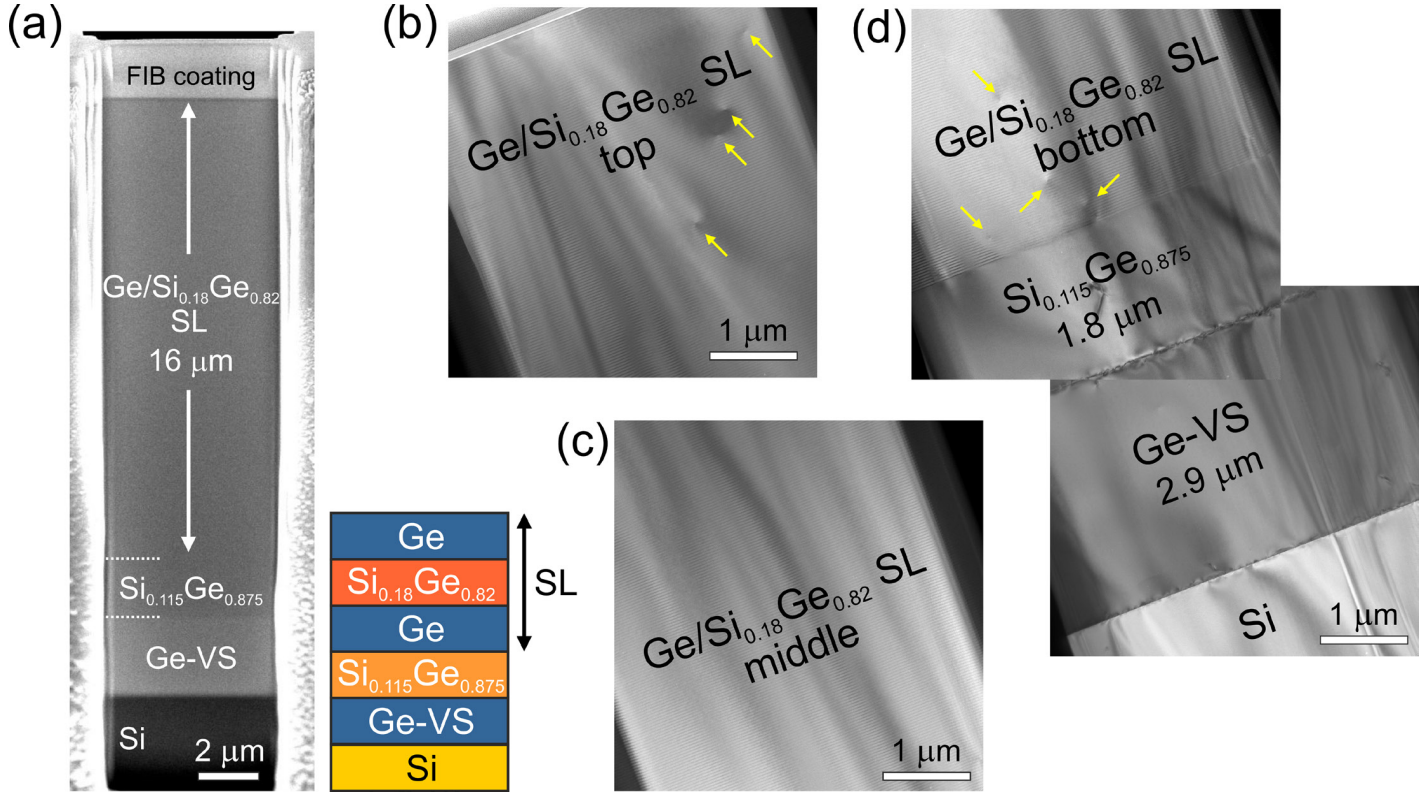

Figure 1

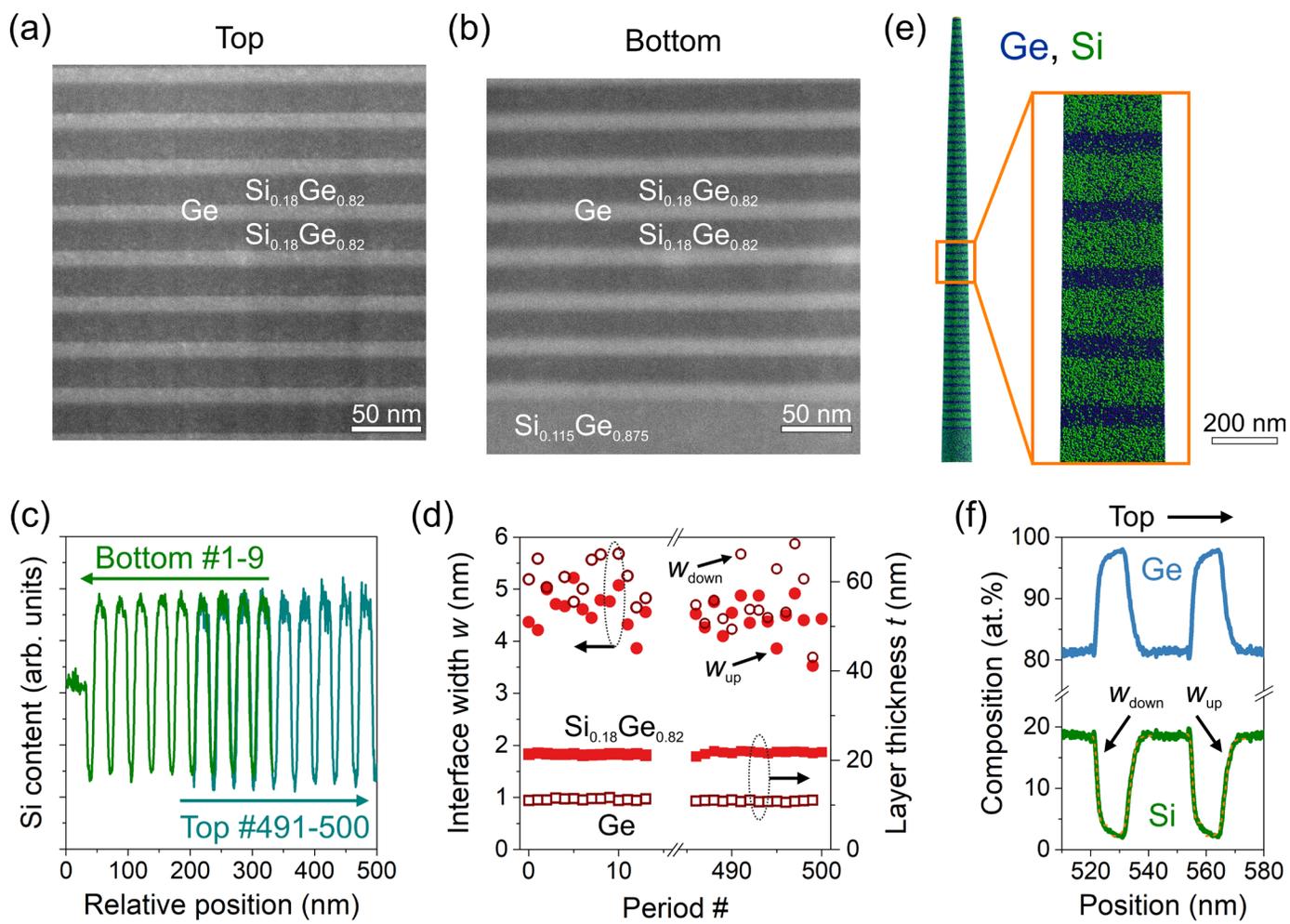

Figure 2

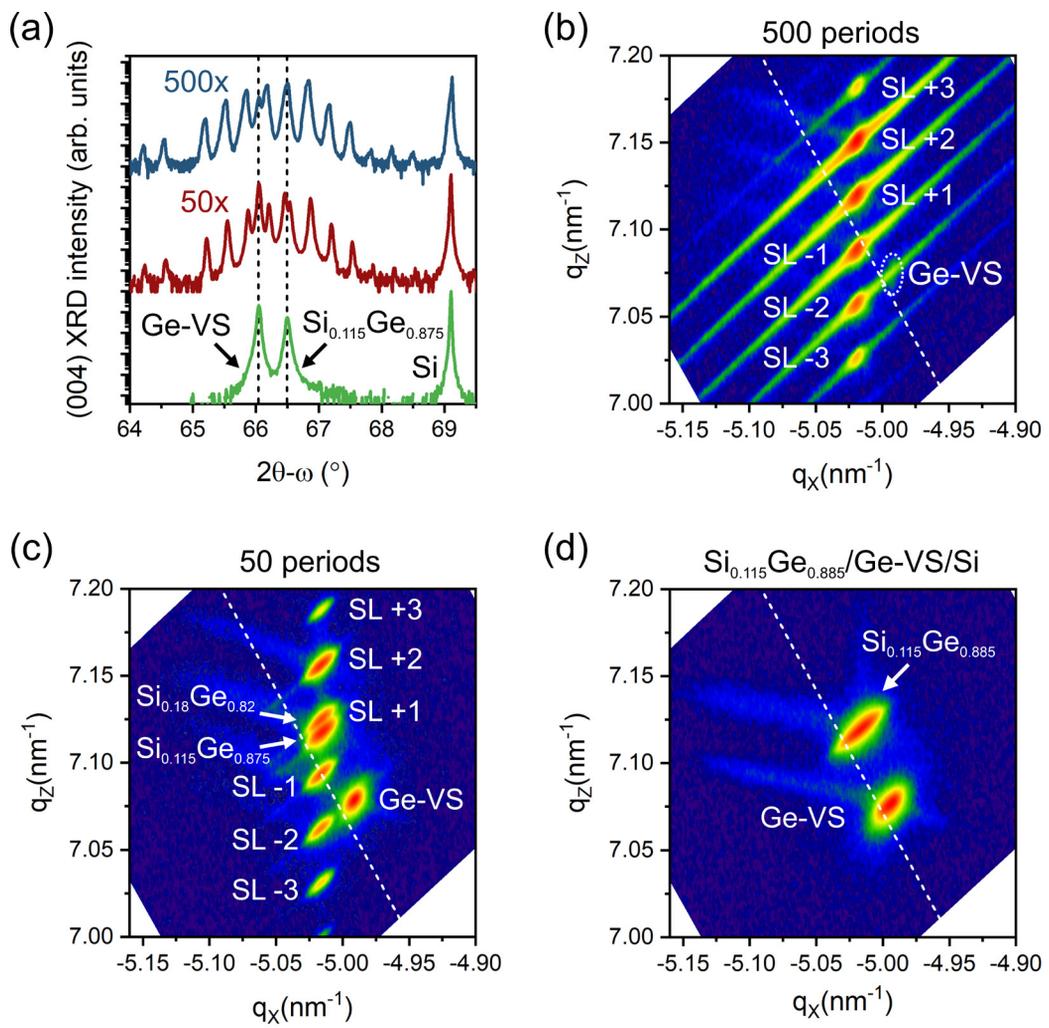

**Figure 3**

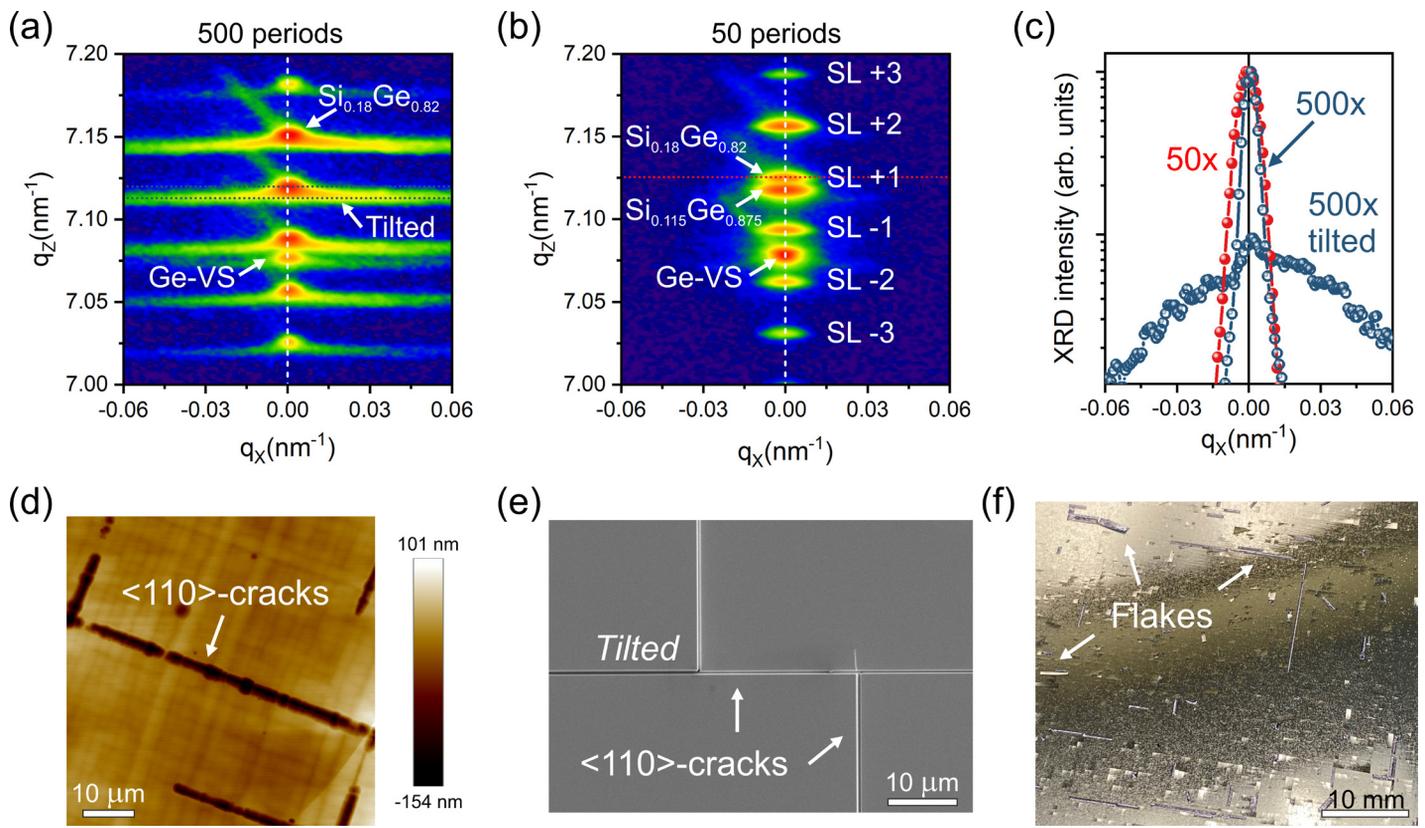

**Figure 4**

# 500-period epitaxial Ge/Si$_{0.18}$Ge$_{0.82}$ multi-quantum wells on silicon

# Supplementary Material

S. Assali,[1,*] S. Koelling,[1] Z. Abboud,[1] J. Nicolas,[1] A. Attiaoui,[1] and O. Moutanabbir[1]

[1] Department of Engineering Physics, École Polytechnique de Montréal, C. P. 6079, Succ. Centre-Ville, Montréal, Québec H3C 3A7, Canada

## Contents

## S1. EDX maps of the 500 periods SL.

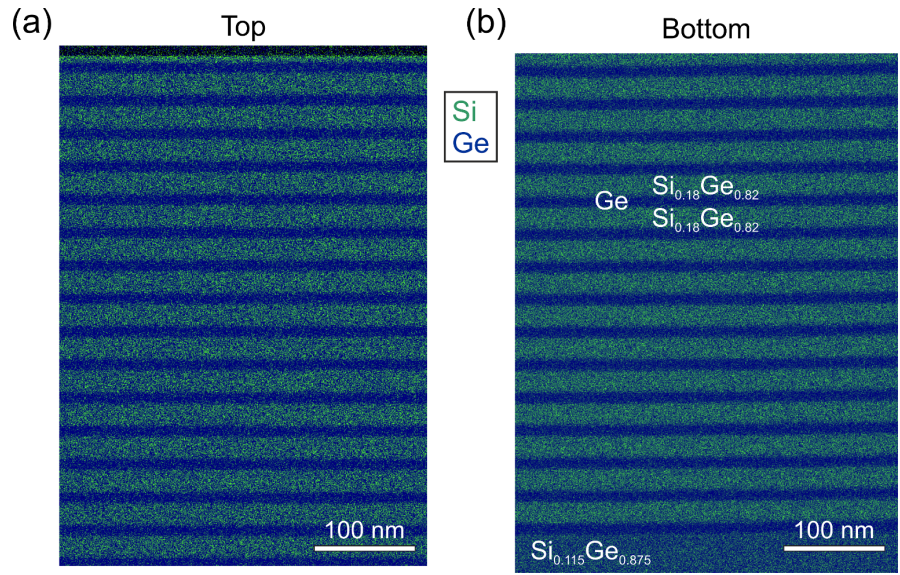

**Figure S1.** (a-b) EDX maps acquired on the top (a) and bottom (b) regions of the 500 period SL.

## S2. STEM on the 50 period SL.

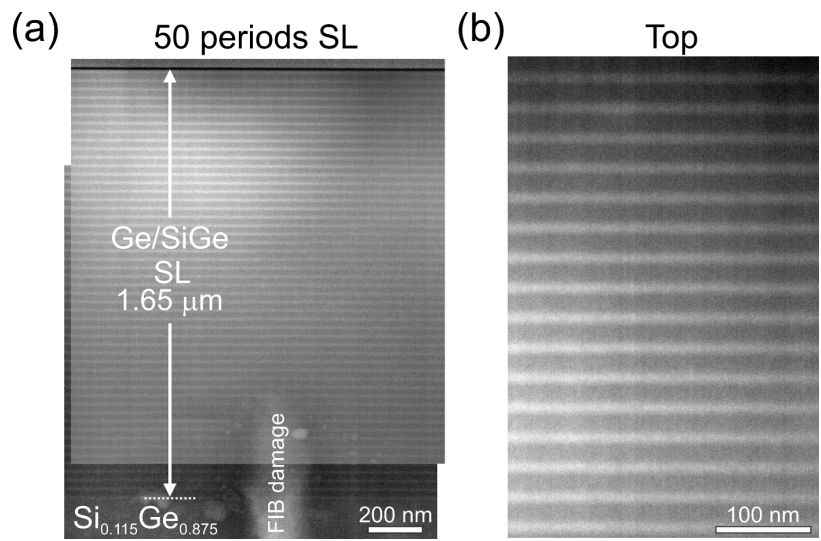

**Figure S2.** (a-b) STEM images of the 50 periods Ge/Si$_{0.18}$Ge$_{0.82}$ SL.

## S3. APT on the 50-period SL.

The Si and Ge compositional profiles on the 50-period estimated from APT are plotted in Fig. S3a, while the thickness $t$ and interface width $w$ fit of the Si profile using Eq. (1) are displayed in Fig. S3b. We note that the reduced volume in the upper portion of the APT tip results in a less precise estimation of the compositional profile close to the sample surface. Consequently, the ($w$, $t$) parameters for the #47-50 layers were omitted (dashed area in Fig. S3c).

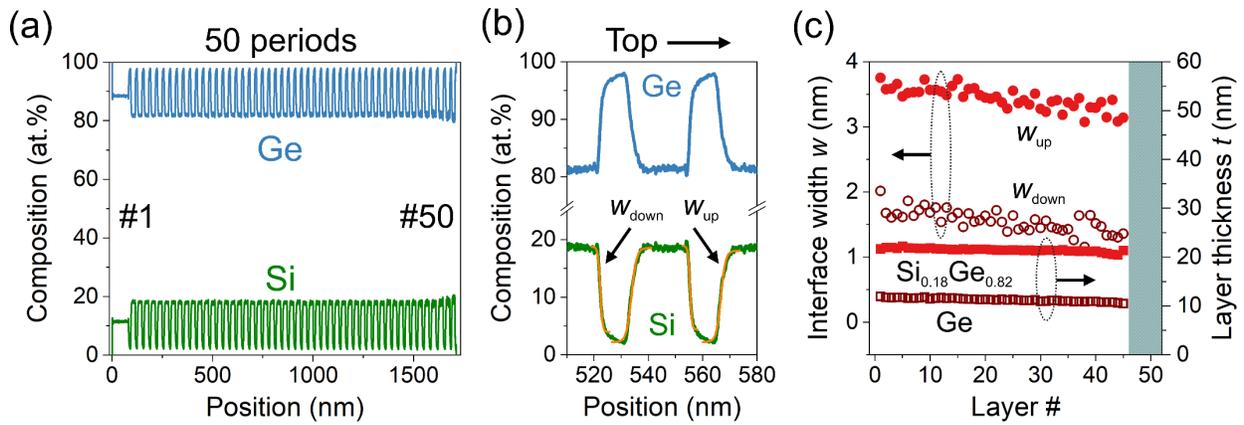

**Figure S3.** (a-b) APT compositional profile for the 50 periods Ge/Si$_{0.18}$Ge$_{0.82}$ SL and fit of the Si profile with Eq. (1). (c) Plot of the thickness $t$ and interface width $w$.

## S4. (004) XRD spectra fitting.

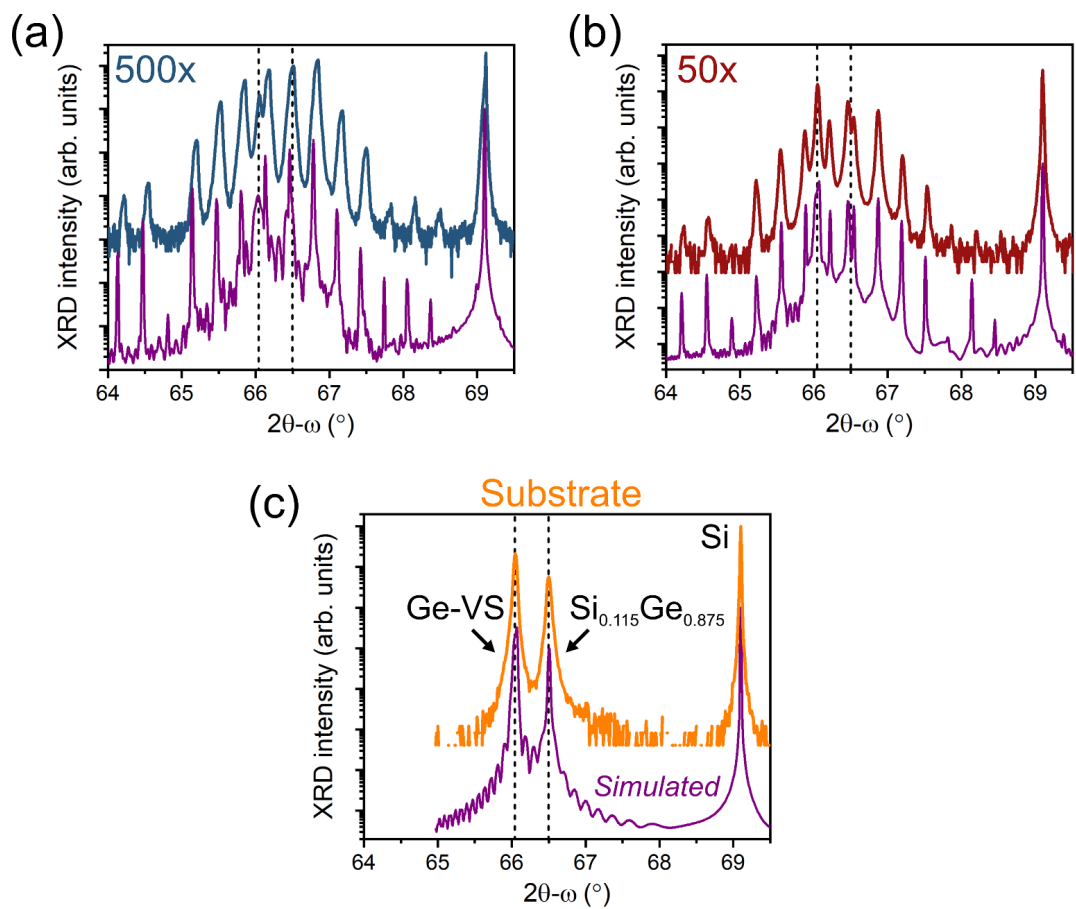

**Figure S4.** (a-c) (004) 2θ-ω XRD curves for the 500 periods SL (a), 50 periods SL (b), and $Si_{0.12}Ge_{0.88}$/Ge-VS/Si (c) samples. The fit of the data using a quasi-kinematic model[1] are shown using purple curves.

## S5. (004) RSM on the Si$_{0.115}$Ge$_{0.875}$/Ge/Si substrate.

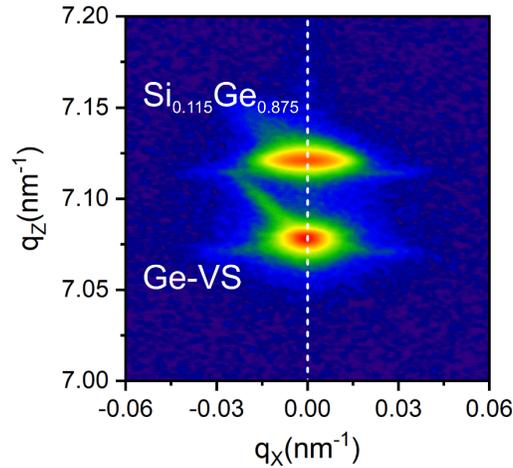

**Figure S5.** RSM around the symmetrical (004) reflection for the Si$_{0.115}$Ge$_{0.875}$/Ge/Si substrate.

## S6. AFM images for all samples.

The evolution of the surface morphology with increasing the SL period is displayed in the 20×20 μm$^2$ AFM images in Fig. S6. The crosshatch morphology is visible and a root mean square (RMS) roughness in the 3-4 nm range is estimated. The $Z_{range} = Z_{max} - Z_{min}$ is in the 10-15 nm range for all SL samples, however the cusping wavelength[2] increases from < 2 nm in Si$_{0.115}$Ge$_{0.875}$/Ge to ~15 μm in the 500 periods SL. The observed surface roughening originates from the strain-driven growth kinetics during lattice-mismatch epitaxy, where mass transport diffusion of atoms at the surface results in the formation of periodic cusps and valleys.[2]

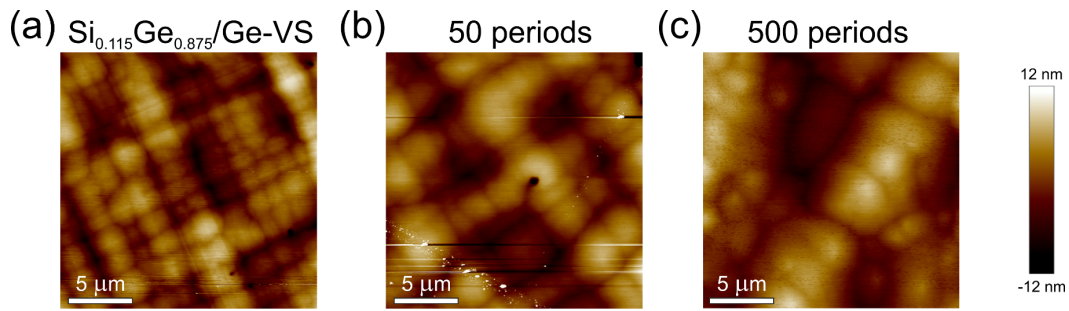

**Figure S6.** 20 μm × 20 μm AFM images for all samples, acquired with the same height scale.

## S7. SEM images of flakes transferred to a Si substrate.

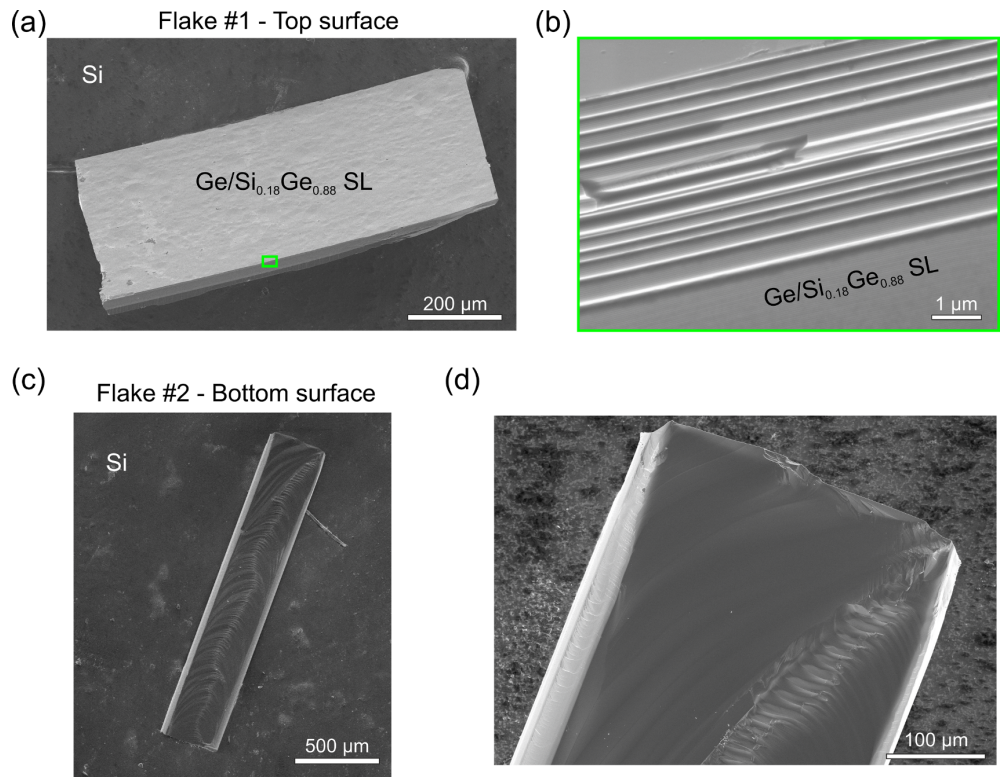

**Figure S7.** (a-d) SEM images of the top (a-b) and bottom (c-d) surface of flakes transferred on a Si substrate.

## References

[1] L. Tapfer and K. Ploog, Physical Review B **40**, 9802 (1989).

[2] H. Gao and W.D. Nix, Annual Review of Materials Science **29**, 173 (1999).